\documentclass[12pt,a4paper]{article}
\usepackage{graphics}
\usepackage{amssymb}
\usepackage{amsbsy}
\usepackage{latexsym}
\usepackage{epsfig}

\newcommand{\be}{\begin{equation}}
\newcommand{\ee}{\end{equation}}
\newcommand{\ba}{\begin{eqnarray}}
\newcommand{\ea}{\end{eqnarray}}

\newcommand{\wbe}{\begin{widetext}}
\newcommand{\wee}{\end{widetext}}

\begin{document}

\begin{titlepage}

\renewcommand{\thefootnote}{\fnsymbol{footnote}}


\vspace{0.3cm}

\begin{center}
{\Large\bf Local free-fall Temperature of modified Schwarzschild
black hole in rainbow spacetime }
\end{center}

\begin{center}
Yong-Wan Kim\footnote{Electronic address:
ywkim65@gmail.com}$^{1}$ and Young-Jai Park$\footnote{Electronic
address: yjpark@sogang.ac.kr}^{2}$\par
\end{center}

\begin{center}

{${}^{1}$Research Institute of Physics and Chemistry,\\ Chonbuk
National University, Jeonju 54896, Korea,}\par {${}^{2}$Department
of Physics, Sogang University, Seoul 04107, Korea}\par
\end{center}

\vskip 0.5cm
\begin{center}
{\today}
\end{center}

\vfill

\begin{abstract}
We obtain a (5+1)-dimensional global flat embedding of modified
Schwarzschild black hole in rainbow gravity. We show that local
free-fall temperature in rainbow gravity, which depends on different
energy $\omega$ of a test particle, is finite at the event horizon
for a freely falling observer, while local temperature is divergent
at the event horizon for a fiducial observer. Moreover, these
temperatures in rainbow gravity satisfy similar relations to those
of the Schwarzschild black hole except overall factor $g(\omega)$,
which plays a key role of rainbow functions in this embedding
approach.
\end{abstract}

\vskip20pt

PACS numbers: 04.70.Dy, 04.20.Jb, 04.62.+v
\vskip15pt
Keywords: Modified Schwarzschild spacetime, Rainbow gravity, Global flat embedding, Unruh effect \\
\end{titlepage}

\newpage

\section{ Introduction}
\setcounter{equation}{0}
\renewcommand{\theequation}{\arabic{section}.\arabic{equation}}

Hawking discovered that a black hole can be described by the
characteristic temperature $T_H$ as seen by an asymptotic
observer~\cite{Hawking:1974sw}. On the other hand, a fiducial
observer staying at a finite distance from a black hole sees thermal
radiation given by the Tolman temperature~\cite{Tolman}.
Furthermore, Unruh~\cite{Unruh:1976db} showed that a uniformly
accelerated observer with a proper acceleration in flat spacetime
detects thermal radiation at the Unruh temperature. These two
effects are, however, closely related each other, {\it i.e.,} the
Hawking effect for a fiducial observer in a curved spacetime can be
considered as the Unruh effect for a uniformly accelerated observer
in a higher dimensional flat spacetime. This provides us a unified
derivation of temperature firstly given by Deser and
Levin~\cite{Deser:1998xb}, and after that there have been much work
on a variety of curved spacetimes~\cite{Kim:2000ct,
Hong:2003xz,Chen:2004qw,Santos:2004ws,Banerjee:2010ma,Cai:2010bv,
Majhi:2011yi,Hu:2011yx,Paston:2014sta,Paston:2014efa} in this line
of the global embedding Minkowskian spacetime (GEMS). Later, this
GEMS approach has been used to define a local temperature for a
freely falling observer outside various black hole
spacetimes~\cite{Brynjolfsson:2008uc,Kim:2009ha,Kim:2013wpa}.

On the other hand, there has also been much attention to modified
dispersion relations (MDR) in gravity's rainbow, which can be viewed
as an extension of doubly special relativity
\cite{AmelinoCamelia:2000mn} in curved spacetime. In particular,
Magueijo and Smolin  \cite{Magueijo:2002am,Magueijo:2002xx} proposed
that the spacetime background felt by a test particle would depend
on its energy $\omega$ such that the energy of the test particle
deforms the background geometry and consequently the dispersion
relation as follows
 \begin{equation}\label{MDR2}
 \omega^2 f^2(\omega/\omega_p)-p^2 g^2(\omega/\omega_p)=m^2,
 \end{equation}
where $p$, $m$, $\omega_p$ are the momentum, the mass of the test
particle, the Planck energy, respectively. Here, the rainbow
functions of $f(\omega/\omega_p) , ~g(\omega/\omega_p)$ with the
conditions $\lim_{\omega \rightarrow 0} f(\omega/\omega_p) = 1$ and~
$\lim_{\omega \rightarrow 0} g(\omega/\omega_p) = 1$ at low energies
are determined depending on the specific models. Since then great
efforts have been devoted to the rainbow gravity related to the
gravity and other stimulated work at the Planck scale
\cite{Liberati:2004ju,Galan:2005ju, Hackett:2005mb,Ling:2006az,
Ling:2006ba,Girelli:2006fw,Ling:2008sy,
Garattini:2011hy,Garattini:2011fs, Amelino-Camelia:2013wha,
Awad:2013nxa,Barrow:2013gia}. In connection with the black hole
thermodynamics, even though the spectrum emitted at infinite from a
black hole would be only marginally affected by the MDR
\cite{Unruh:1980cg}, there have been much work on black hole physics
in the rainbow gravity \cite{AmelinoCamelia:2005ik, Ling:2005bp,
 Liu:2007fk, Peng:2007nj, Li:2008gs, Garattini:2009nq}.
In fact, the temperature as well as the entropy of a black hole
probed by a test particle receives energy dependent corrections due
to the modification of the MDR. Recently, the black hole
thermodynamics in the rainbow gravity was studied with the following
choice of rainbow functions
\cite{Ali:2014xqa,Gim:2014ira,Ali:2014qra,Gim:2015zra} among many
others
\cite{Magueijo:2002am,Garattini:2011hy,Awad:2013nxa,Ling:2005bp,
Liu:2007fk,Peng:2007nj,Garattini:2009nq},
\begin{equation}\label{rainbowfunc}
f(\omega/\omega_p)=1, \quad g(\omega/\omega_p)=\sqrt{1-\eta
\left({\omega}/{\omega_p}\right)^n},
\end{equation}
where $n$ is a positive integer and $\eta$ is a constant of order
unity. This was originally motivated from a quantum spacetime
phenomenology \cite{AmelinoCamelia:2008qg}, and is compatible with
some of the results in Loop Quantum Gravity \cite{Gambini:1998it}
and $\kappa$-Minkowski noncommutative spacetime
\cite{Lukierski:1993wx}.
From now on, we will take
$\omega_p=1$, for simplicity, unless mentioned otherwise.


In this paper, we wish to study the GEMS of the modified
Schwarzschild black hole in the rainbow gravity. We show that the
energy dependent local free-fall temperature is finite at the event
horizon for a freely falling observer, while the energy dependent
local temperature is divergent for a fiducial observer. In section
2, we briefly review the structure of the GEMS of the Schwarzschild
spacetime and its local free-fall temperature. In section 3, we
apply this GEMS approach to the modified Schwarzschild black hole in
the rainbow gravity, and find their corresponding local free-fall
temperature. As a result, we find that the temperatures in the
modified Schwarzschild black hole keeps the same as those in the
Schwarzschild black hole, except overall factor of $g(\omega)$. Our
conclusions are drawn in section 4.

\section{GEMS of the Schwarzschild black hole}
\setcounter{equation}{0}
\renewcommand{\theequation}{\arabic{section}.\arabic{equation}}

First, let us briefly recapitulate the GEMS embedding of the
Schwarzschild spacetime with the metric
 \be
 (ds_{4})^2=-h(r)dt^2+h^{-1}(r)dr^2
      +r^2(d\theta^2+\sin^2\theta d\phi^2),
 \ee
which can be embedded into a (5+1)-dimensional Minkowskian spacetime
 of
 \be
 (ds_{6})^2=\eta_{IJ}dz^I dz^J
 \ee
with $h(r)=1-2M/r$ and a metric $\eta_{IJ}={\rm
diag}(-1,1,1,1,1,1)$. The two metrics are related through the
embedding functions $z^I=z^I(x^\mu)$ with the metric
$g_{\mu\nu}=\eta_{IJ}{{\partial z^I}/{\partial x^\mu}}{{\partial
z^J}/{\partial x^\nu}}$. Then, $z^I(x^\mu)$ \cite{Fronsdal:1959zza}
are explicitly given by
 \ba\label{schemb1}
 z^0&=&{k^{-1}_H}{\sqrt{h(r)}}\sinh(k_H t),\\
 z^1&=&{k^{-1}_H}{\sqrt{h(r)}}\cosh(k_H t),\\
 z^2&=&\int dr \sqrt{\frac{2M(r^2+2Mr+4M^2)}{r^3}},\\
 z^3&=&r\sin\theta\cos\phi,\\
 z^4&=&r\sin\theta\sin\phi,\\
 z^5&=&r\cos\theta.
 \ea
Here, we denote the surface gravity $k_H$ on the event horizon of
the Schwarzschild black hole as
\begin{equation}\label{sg-Sch}
k_H = -\frac{1}{2}\lim_{r\rightarrow r_H}
{\sqrt{\frac{-g^{11}}{g^{00}}}}\frac{(g^{00})'}{g^{00}}=\frac{1}{2r_H},
\end{equation}
where the event horizon $r_H$ is given by $2M$.

Then, a uniformly accelerating observer follows a hyperbolic
trajectory in the (5+1)-dimensional flat spacetime described by a
proper acceleration
 \be\label{Sch-accel}
 a^{-2}_6=(z^1)^2-(z^0)^2=16M^2h(r).
 \ee
Thus, the Unruh temperature~\cite{Unruh:1976db} can be easily read
as
 \be\label{unruh-sch}
 T_U=\frac{a_6}{2\pi}=\frac{1}{8\pi M\sqrt{h(r)}}.
 \ee
This corresponds to the local temperature measured by a fiducial
observer staying at a finite distance from the black hole, the
so-called fiducial temperature
 \be\label{fid-temp}
 T_{\rm FID}=\frac{T_H}{\sqrt{-g_{00}}},
 \ee
where the Hawking temperature $T_H$ is measured by an asymptotic
observer
 \be
 T_H=\frac{1}{8\pi M}.
 \ee
As a result, one can say that the Hawking effect for a fiducial
observer in the black hole spacetime is just the Unruh effect for a
uniformly accelerated observer in a higher-dimensional flat
spacetime.

Next, let us consider a freely falling observer who has been dropped
from rest at $r=r_0$ and at $\tau=0$. For a freely falling observer,
there are turning
points~\cite{Brynjolfsson:2008uc,Kim:2009ha,Kim:2013wpa} of radial
geodesics where a freely falling observer is momentarily at rest
with respect to black holes.

For a Killing vector of $K^{\mu}=(\partial_t)^{\mu}=(1,0,0,0)$ related to the energy conservation,
we have a constant of motion
 \be\label{kil}
 K_\mu\frac{dx^\mu}{d\lambda}=-h(r)
     \frac{dx^\mu}{d\lambda}\equiv-E.
 \ee
Then, timelike equation
 \be
 1=-g_{\mu\nu}\frac{dx^\mu}{d\lambda}\frac{dx^\nu}{d\lambda}
 \ee
 becomes
 \be\label{teq}
 1=h(r)\left(\frac{dt}{d\lambda}\right)^2
   -h^{-1}(r){\left(\frac{dr}{d\lambda}\right)^2}
 \ee
along with an equatorial plane $\theta=\pi/2$. From Eqs. (\ref{kil})
and (\ref{teq}),  one can find the following orbit
equations~\cite{Brynjolfsson:2008uc,Kim:2009ha,Kim:2013wpa}
 \ba\label{orbiteq}
 \frac{dt}{d\tau}&=&\frac{\sqrt{h(r_0)}}{h(r)},\\
 \frac{dr}{d\tau}&=&-\sqrt{h(r_0)-h(r)}.
 \ea
\begin{figure*}[t!]
   \centering
   \includegraphics{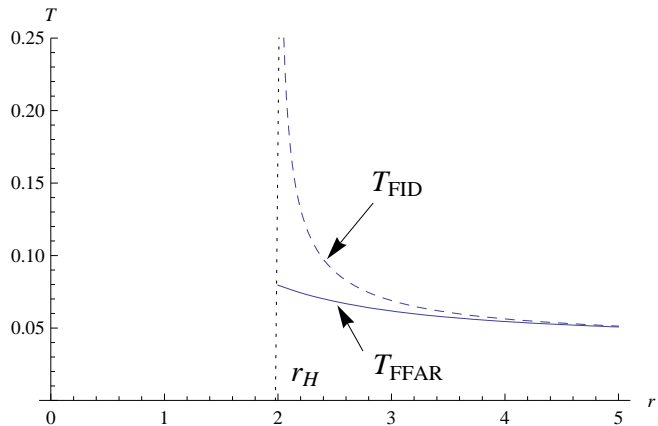}
\caption{The local temperatures $T_{FFAR}$ and $T_{FID}$ plotted as
a function of the radial variable $r$. The free-fall temperature
$T_{FFAR}$ seen by a freely falling observer remains finite at the
horizon, while the fiducial temperature  $T_{FID}$ seen by an
observer staying at a finite distance from the horizon blows up.}
 \label{fig:fig_1}
\end{figure*}
Therefore, one can obtain the $\widetilde{a}_6$ acceleration at the
turning point $r=r_0$ given by
 \be\label{a6-sch}
 (\widetilde{a}_6)^2~=~\eta_{IJ} \widetilde{a}^I \widetilde{a}^J~=~\frac{r^3+2Mr^2+4M^2r+8M^3}{16M^2r^3}.
 \ee
Now, the local free-fall temperature measured by the freely falling
observer at rest (FFAR) can be obtained as
 \be\label{ffar-sch}
 T_{\rm
 FFAR}=\frac{\widetilde{a}_6}{2\pi}=\sqrt{\sum^3_{n=0}\left(\frac{2M}{r}\right)^n}T_H.
 \ee
In Figure~\ref{fig:fig_1} we plot the two temperatures $T_{FID}$ and
$T_{FFAR}$ as functions of the radial variable $r$. Note that as
shown in Figure 1, the local free-fall temperature at the event
horizon is finite as $T_{\rm FFAR}=2T_H$, while the local fiducial
temperature $T_{\rm FID}$ (\ref{fid-temp}) for the fiducial observer
diverges as $r\rightarrow r_H$ as shown by Brynjolfsson and
Thorlacius.~\cite{Brynjolfsson:2008uc}

\section{Modified Schwarzschild black hole in rainbow spacetime}
\setcounter{equation}{0}
\renewcommand{\theequation}{\arabic{section}.\arabic{equation}}

Now, we are ready to study the modified Schwarzschild black hole in the rainbow spacetime,
which is described by the metric
 \be
 (ds^{\omega}_4)^2=-\frac{1}{f^2(\omega)}h(r)dt^2
      +\frac{1}{g^2(\omega)}h^{-1}(r)dr^2
      +\frac{r^2}{g^2(\omega)}(d\theta^2+\sin^2\theta d\phi^2).
 \ee
This spacetime can be embedded into a (5+1)-dimensional Minkowskian
spacetime
 \be
 (ds^{\omega}_6)^2=\eta_{IJ}dz'^I dz'^J
 \ee
with a metric $\eta_{IJ}={\rm diag}(-1,1,1,1,1,1)$. Explicitly, the
embedding functions $z'^I(x^\mu)$ are given by the following
coordinate transformations depending on different energy $\omega$ of
the probe \ba
 z'^0&=&\frac{\sqrt{h(r)}}{k^\omega_Hf(\omega)}\sinh(k^\omega_H t),\label{Rschemb1}\\
 z'^1&=&\frac{\sqrt{h(r)}}{k^\omega_Hf(\omega)}\cosh(k^\omega_H t),\label{Rschemb2}\\
 z'^2&=&\frac{1}{g(\omega)}\int dr \sqrt{\frac{2M (r^2 +2Mr +4M^2)}{r^3}},\\
 z'^3&=&\frac{1}{g(\omega)}r\sin\theta\cos\phi,\\
 z'^4&=&\frac{1}{g(\omega)}r\sin\theta\sin\phi,\\
 z'^5&=&\frac{1}{g(\omega)}r\cos\theta.
\ea
Note here that the surface gravity $k^{\omega}_H$ on the event
horizon in the rainbow gravity, is defined by
\begin{equation}
k^{\omega}_H = \frac{g(\omega)}{f(\omega)}k_H.
\end{equation}
Then, the Hawking temperature $T^{\omega}_H$ measured by an
asymptotic observer can be written
 \be\label{T2}
 T^{\omega}_H=\frac{k^{\omega}_H}{2
 \pi}=\frac{g(\omega)}{f(\omega)}T_H.
 \ee
Note that Eq. $(\ref{T2})$ indicates that the temperature of the
modified Schwarzschild black hole is different for a probe with
different energy $\omega$.

It seems appropriate to comment that the factors
$k^{\omega}_Hf(\omega) $ in Eqs. (\ref{Rschemb1}) and
(\ref{Rschemb2}) can be rewritten as $g(\omega)k_H$. Then, we have
the relation between the line elements $ds^{\omega}_6$ and $ds_6$
as \be
 (ds^\omega_6)^2={g^{-2}(\omega)}(ds_6)^2.
 \ee
Thus, it seems particularly that only the $g(\omega)$ plays the key
role of the two rainbow functions in the GEMS embedding.

On the other hand, an observer, who is uniformly accelerated in the
(5+1)-dimensional flat spacetime, follows a hyperbolic trajectory
described by a proper acceleration
 \be\label{rSch-accel}
 (a^{\omega}_6)^{-2}=(z'^1)^2-(z'^0)^2={g^{-2}(\omega)}(a_6)^{-2}.
 \ee
Then, the Unruh temperature can be read as
 \be\label{unruh-rsch}
 T^{\omega}_U=\frac{a^{\omega}_6}{2\pi}={g(\omega)}T_U.
 \ee
In fact, this corresponds to the fiducial temperature measured by a
fiducial observer at a finite distance from the black hole
 \be\label{rfid-temp}
 T^{\omega}_{\rm FID}=\frac{T^{\omega}_H}{\sqrt{-g^{\omega}_{00}}}=g(\omega) T_{\rm FID}.
 \ee
As a result, the Hawking effect for a fiducial observer in the
rainbow spacetime can also be said to be the Unruh effect for a
uniformly accelerated observer in a higher-dimensional flat
spacetime as like in the Schwarzschild spacetime. Moreover, both of
them are only proportional to the rainbow function $g(\omega)$ in
common.


Now, consider a freely falling observer who is dropped from rest
at $r=r_0$ and $\tau=0$. Making use of a constant of motion as in
Eq. (\ref{kil}) for the modified Schwarzschild spacetime in the
rainbow gravity
 \ba
 \frac{dt}{d\tau}&=& \frac{f(\omega)\sqrt{h(r_0)}}{h(r)},\nonumber\\
 \frac{dr}{d\tau}&=& -g(\omega)\sqrt{h(r_0)-h(r)},
 \ea
one can find the acceleration $\widetilde{a}^{\omega}_6$ for the
freely falling observer
 \be
 (\widetilde{a}^{\omega}_6)^2=g^2(\omega)(\widetilde{a}_6)^2,
 \ee
where $\widetilde{a}_6$ is the acceleration given by (\ref{a6-sch}).
Therefore, the local free-fall temperature seen by a freely falling
observer at rest can be written by
 \be\label{ffar-sch1}
 T^{\omega}_{\rm FFAR}=\frac{\widetilde{a}^{\omega}_6}{2\pi}
  =g(\omega)\sqrt{\sum^3_{n=0}\left(\frac{2M}{r}\right)^n}T_H={g(\omega)}{T}_{\rm FFAR}.
 \ee
\begin{figure*}[t!]
   \centering
   \includegraphics{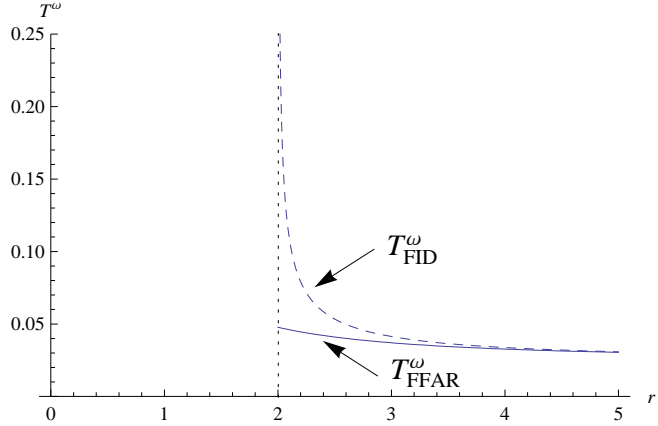}
\caption{The local temperatures $T^\omega_{FFAR}$ and
$T^\omega_{FID}$ plotted as functions of the radial variable $r$
with $\eta=1$, $\omega=0.8$, and $n=2$. Similar to the Schwarzschild
case, the free-fall temperature $T^\omega_{FFAR}$ also remains
finite at the horizon while the fiducial temperature
$T^\omega_{FID}$ blows up.}
 \label{fig:fig_2}
\end{figure*}
\begin{figure*}[t!]
   \centering
   \includegraphics{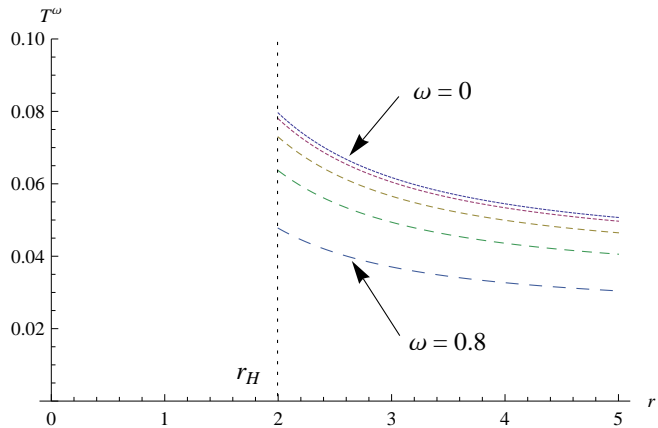}
\caption{The local free-fall temperature ${T}^{\omega}_{FFAR}$
plotted as functions of the radial variable $r$ with $\eta=1$,
$n=2$, and $\omega=0$, $0.2$, $0.4$, $0.6$, $0.8$, respectively. The
curve with $\omega=0$ corresponds to $T_{FFAR}$ in Eq.
(\ref{ffar-sch}). The curves show that the local free-fall
temperatures become finite as $r\rightarrow r_H$ regardless of the
energy $\omega$ of a test particle.}
 \label{fig:fig_3}
\end{figure*}
In Figure~\ref{fig:fig_2}, we plot the two temperatures
$T^\omega_{FFAR}$ and $T^\omega_{FID}$ as functions of the radial
variable $r$ and a fixed $\omega=0.8$. It is important to note that
similar to the Schwarzschild case, the local free-fall temperature
at the event horizon is finite as $T^{\omega}_{\rm
FFAR}=2{g(\omega)}T_H$, while the local temperature $T^\omega_{\rm
FID}$ for the fiducial observer diverges as $r\rightarrow r_H$. In
Figure \ref{fig:fig_3} and \ref{fig:fig_4}, we also plot the local
temperatures of ${T}^{\omega}_{FFAR}$ and ${T}^{\omega}_{FID}$,
respectively, as functions of the radial variable $r$ and $\omega$
with $\eta=1$, $n=2$, which show that the local free-fall
temperature ${T}^{\omega}_{FFAR}$ is less sensitive to the energy
$\omega$ of the probe.
\begin{figure*}[t!]
   \centering
   \includegraphics{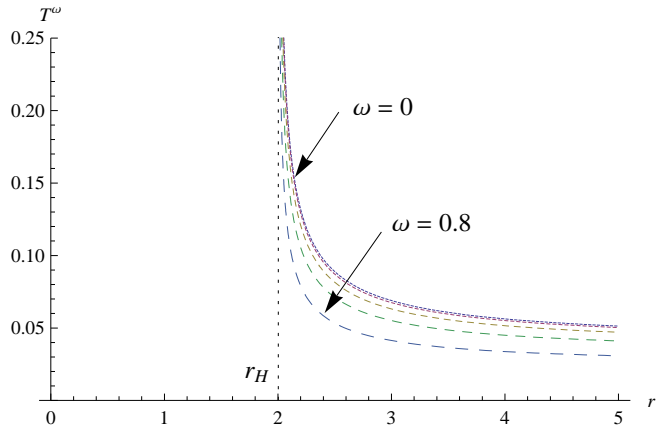}
\caption{The local fiducial temperature ${T}^{\omega}_{FID}$ plotted
as functions of the radial variable $r$ with $\eta=1$, $n=2$, and
$\omega=0$, $0.2$, $0.4$, $0.6$, $0.8$, respectively. The curve with
$\omega=0$ corresponds to $T_{FID}$ in Eq. (\ref{fid-temp}). They
show that the local fiducial temperatures diverge as $r\rightarrow
r_H$  regardless of the energy $\omega$ of a test particle.}
 \label{fig:fig_4}
\end{figure*}
\section{Conclusions}
\setcounter{equation}{0}
\renewcommand{\theequation}{\arabic{section}.\arabic{equation}}
\label{sec:conclusions}

In this paper, we have obtained a (5+1)-dimensional global flat
embedding of the modified Schwarzschild black hole in rainbow
gravity. We have shown that with the proper choice of the rainbow
functions (\ref{rainbowfunc}), the local free-fall temperature for a
freely falling observer is finite at the event horizon, while the
local temperature for a fiducial observer is divergent, which are
similar to the case of the Schwarzschild black hole, except that
these temperatures also depend on different energy $\omega$ of a
test particle.

On the other hand, it seems to be important to note that in the GEMS
approach all the local temperatures of $T^\omega_U$,
$T^\omega_{FID}$, and $T^\omega_{FFAR}$ depend only on the
$g(\omega)$ of the two rainbow functions in Eq. (\ref{rainbowfunc}).


\end{document}